\documentclass[conference]{IEEEtran}

\usepackage[T1]{fontenc}% optional T1 font encoding

\usepackage{graphicx}
\usepackage[noadjust]{cite}
\usepackage{mcite}
\usepackage{amsfonts,helvet}
\usepackage{fancyhdr}
\usepackage{threeparttable}
\usepackage{epsf,epsfig}
\usepackage{amsthm}
\usepackage{amsmath}
\usepackage{siunitx}
\usepackage{amssymb}
\usepackage{dsfont}
\usepackage{subfigure}
\usepackage{color}
\usepackage{enumerate}
\usepackage[bookmarks=false]{hyperref}
\usepackage{cancel}
\usepackage{bbm}
\usepackage{dsfont}
\usepackage[subnum]{cases}
\usepackage{adjustbox}
\usepackage[linesnumbered,ruled,noend]{algorithm2e}
\usepackage{multicol}
\usepackage[english]{babel}
\usepackage{enumitem}
\usepackage{float}

\newtheorem{theorem}{Theorem}

\newtheorem{corollary}{Corollary}

\newmuskip\pFqmuskip

\def\bff{{\bf f}}

\def\bh{{\bf h}}

\def\bn{{\bf n}}

\def\bq{{\bf q}}
\def\br{{\bf r}}
\def\bs{{\bf s}}

\def\bw{{\bf w}}

\def\bC{{\bf C}}
\def\bD{{\bf D}}

\def\bF{{\bf F}}

\def\bH{{\bf H}}
\def\bI{{\bf I}}

\def\bK{{\bf K}}

\def\bM{{\bf M}}

\def\bW{{\bf W}}

\def\cP{\mbox{$\mathcal{P}$}}
\def\cQ{\mbox{$\mathcal{Q}$}}

\def\bbC{\mbox{$\mathbb{C}$}}

\def\bbR{\mbox{$\mathbb{R}$}}

\def\btau{\boldsymbol{\tau}}
\def\bSigma{\boldsymbol{\Sigma}}

\usepackage{array}

\makeatletter
\newcommand{\thickhline}{%
    \noalign {\ifnum 0=`}\fi \hrule height 1pt
    \futurelet \reserved@a \@xhline
}
\newcolumntype{"}{@{\hskip\tabcolsep\vrule width 1pt\hskip\tabcolsep}}
\makeatother

\IEEEoverridecommandlockouts

\title{\huge
Coordinated Multicell Beamforming and Power Allocation 
\\ for Massive MIMO with Low-Resolution ADC/DAC
}

\author{Yunseong Cho, Jinseok Choi$^\dagger$, and Brian L. Evans \\
\IEEEauthorblockA{\normalsize{Wireless Networking and Communications Group, The University of Texas at Austin}\\
Email: yscho@utexas.edu, bevans@ece.utexas.edu\\
$^\dagger$Ulsan National Institute of Science and Technology (UNIST), Ulsan, Republic of Korea \\
Email: jinseokchoi@unist.ac.kr}
}

\begin{document}
\maketitle

\begin{abstract}

In this work, we present a solution for coordinated beamforming and power allocation when base stations employ a massive number of antennas equipped with low-resolution analog-to-digital and digital-to-analog converters. We address total power minimization problems of the coarsely quantized uplink (UL) and downlink (DL) communication systems with target signal-to-interference-plus-noise ratio (SINR) constraints. By combining the UL problem with minimum mean square error combiners and deriving the Lagrangian dual of the DL problem, we prove UL-DL duality and show there is no duality gap even with coarse data converters. Inspired by strong duality, we devise an iterative algorithm to determine the optimal UL transmit powers, and then linearly amplify the UL combiners with proper weights to acquire the optimal DL precoder. Simulation results validate strong duality and evaluate the proposed method in terms of total power consumption and achieved SINR.

%%%%%%%%%%%%%%%%%%%%%%%%%%%%
\end{abstract}
%%%%%%%%%%%%%%%%%%%%%%%%%%%%

%%%%%%%%%%%%%%%%%%%%%%%%%%%
\section{Introduction}
\label{sec:intro}
%%%%%%%%%%%%%%%%%%%%%%%%%%%

% BACKGROUND & MOTIVATION 
% 1) Massive MIMO 
% 2) Low-resolution 
% 3) CoMP motivation
Massive multiple-input-multiple-output (MIMO) has been considered as a key technique for next-generation communication systems because of its advantage in spectral efficiency \cite{marzetta2010noncooperative}.
Associating each antenna with power-hungry high-resolution analog-to-digital converters (ADCs) and digital-to-analog converters (DACs) would consume considerable power and become a main challenge for realistic deployment.
Consequently, transceivers with low-resolution data converters have been gathering momentum \cite{choi2016near,studer2016quantized,choi2019robust,cho2019one,choi2017resolution,orhan2015low, xu2019uplink, choi2019two,jacobsson2017quantized, dai2019achievable}. 
Moreover, in multicell systems, the in-cell channel and out-of-cell interferers as well as the non-negligible quantization error should be modeled and mitigated in the design of the communication system.

% ChEst & Detection
State-of-the-art data detector and channel estimation have been developed for low-resolution ADCs \cite{choi2016near, studer2016quantized, choi2019robust,cho2019one}.
The authors in \cite{choi2019robust} proposed a learning-based detector with an artificial noise to overcome an issue of 1-bit ADCs, stochastic resonance.
In \cite{cho2019one}, a coding-theoretic approach was given to perform soft detection and its refinement under 1-bit ADCs.
% resolution-adaptive-ADC
Resolution-adaptive ADCs design and the corresponding ADC bit-allocation algorithm were derived in \cite{choi2017resolution}.
% AQNM
For tractability, an additive quantization noise model (AQNM) was used in \cite{orhan2015low, xu2019uplink, choi2019two} with informative analyses.
% Low-resolution DAC %
Low-resolution DAC systems have also been investigated \cite{jacobsson2017quantized,dai2019achievable}.
% General Few-bit & 1-bit
In \cite{jacobsson2017quantized}, it was shown that achievable rates of 3-4 bits DACs are comparable to infinite-resolution DACs.
The AQNM was also utilized in \cite{dai2019achievable} to approximate the uplink (UL) and downlink (DL) achievable rates in full-duplex systems.

\begin{figure}[!t]\centering
\includegraphics[width=0.95\columnwidth]{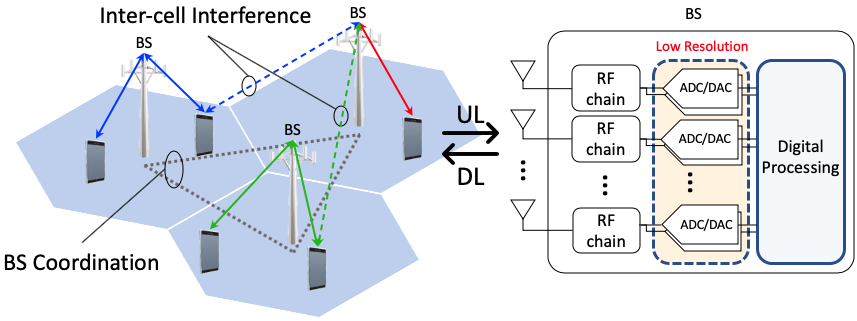}
\vspace{-1.0em}
\caption{Multicell multiuser-MIMO configuration with low-resolution ADCs and DACs at the base station (BS).} 
\label{fig:system}
\vspace{-1.0em}
\end{figure}

% part 1) CoMP literature1
As one of the key ingredients of modern cellular systems, a coordinated multipoint (CoMP) design across base stations (BSs) has shown a significant gain in communication performance \cite{rashid1998joint,rashid1998transmit,dahrouj2010coordinated,jungnickel2014role}.
In \cite{rashid1998joint}, beamforming (BF) and power allocation (PA) in UL CoMP were developed by using a fixed-point iteration method.
Considering DL as a virtual UL, UL-DL CoMP BF and PA were further proposed in \cite{rashid1998transmit} using only local measurement in a distributed fashion.
The authors in \cite{dahrouj2010coordinated} further attained Lagrangian-based duality for multiuser MIMO systems and proposed a distributed method with less complexity load on users and BSs.
A vision on the possible combinations of massive MIMO and CoMP architectures was described in \cite{jungnickel2014role} achieving higher throughput.

% Contribution
In this work, we integrate the low-resolution converters into the CoMP BF and PA designs. 
Under the non-negligible quantizer errors that must be carefully regarded, we first write the UL and DL problems whose purposes are to minimize total transmit power with individual signal-to-interference-plus-noise ratio (SINR) constraints. 
We then prove that UL-DL duality holds under coarse quantizers by showing that the UL problem processed by the minimum mean square error (MMSE) combiner is equal to the Lagrangian dual of the DL problem.
We show that there is zero duality gap by converting the DL problem to a strictly feasible second-order cone program.
Based on strong UL-DL duality, we devise an iterative algorithm to solve the UL problem in a distributed manner with convergence to an optimum.
We also state that an optimal DL BF can be a mixture of the UL combiner and weights computed by the UL result.
Numerical results show that the proposed design outperforms a conventional method in terms of total power and achievable SINR.

%%% NOTATIONS %%%
{\it Notation}: $\bf{A}$ is a matrix and $\bf{a}$ is a column vector. 
$\mathbf{A}^{H}$ and $\mathbf{A}^T$  denote conjugate transpose and transpose. 
$[{\bf A}]_{i,:}$ and $ \mathbf{a}_i$ are the $i$th row and column vectors of $\bf A$. 
We denote $a_{i,j}$ as the $\{i,j\}$th element of $\bf A$ and $a_{i}$ as the $i$th element of $\bf a$. 
$\mathcal{CN}(\mu, \sigma^2)$ is a complex Gaussian distribution with mean $\mu$ and variance $\sigma^2$. 
The diagonal matrix $\rm diag(\bf A)$ has $\{a_{i,i}\}$ as its diagonal entries, and $\rm diag (\bf a)$ has $\{a_i\}$ as its diagonal entries. 
A block diagonal matrix is presented as ${\rm blkdiag}({\bf A}_1, \dots,{\bf A}_{N})$. 
$\|\bf A\|$ is L2 norm. 
${\bf I}_N$ is a $N\times N$ identity matrix and ${\bf 0}_N$ is a $N \times 1$ zero vector.

%%%%%%%%%%%%%%%%%%%%%%%%%%%%
\section{System Model}
\label{sec:sys_model}
%%%%%%%%%%%%%%%%%%%%%%%%%%%%

% %FIGURE 
% \begin{figure}[!t]\centering
% \includegraphics[scale = 0.7]{massiveMIMomulticell.png}
% \caption{\color{blue} Massive MIMO multicell network with coordinated beamforming.} 
% \label{fig:system}
% \end{figure}

We consider a multicell multiuser-MIMO network with $N_c$ cells, $N_{u}$ single-antenna users per cell as shown in Fig.~\ref{fig:system}, with time division multiplexing (TDD) assumption.
Users in cell $i$ mainly communicate with a designated BS in cell $i$ (BS$_i$) equipped with $N_b$ antennas.
We assume that the BSs have low-resolution ADCs and DACs with the same $b$-bits precision and the users have infinite-resolution quantizers.

%%%%%%%%%%%%
\subsection{Uplink System}
%%%%%%%%%%%

% We assume large antenna arrays followed by low-resolution ADCs at the BS. 
Each user $u$ in cell $i$ transmits signal $x^{\rm ul}_{i,u} \!=\! \sqrt{\lambda_{i,u}}s^{\rm ul}_{i,u}$ where $\lambda_{i,u}$ and $s^{\rm ul}_{i,u}$ are transmit power and a symbol, respectively. 
The channel between user $u$ in cell $j$ and the BS$_i$ is ${\bf h}_{i,j,u} \in \bbC^{N_b}$.
The received signal at BS$_i$ is expressed as
\begin{align}
    \label{eq:r_ul}
    {\bf r}^{\rm ul}_i 
    &={\bf H}_{i,i}{\bf x}^{\rm ul}_i + {\textstyle\sum}_{j\neq i}^{N_c}{\bf H}_{i,j}{\bf x}^{\rm ul}_j + {\bf n}^{\rm ul}_i \nonumber \\
    &={\bf H}_{i,i}{\pmb \Lambda}^{1/2}_i{\bf s}^{\rm ul}_i + {\textstyle\sum}_{j\neq i}^{N_c}{\bf H}_{i,j}{\pmb \Lambda}^{1/2}_j{\bf s}^{\rm ul}_j + {\bf n}^{\rm ul}_i
\end{align} 
where ${\bf H}_{i,j} \!\in\! \bbC^{N_b\times N_u}$ is the channel between BS$_i$ and users in  cell $j$, whose $u$th column is $\bh_{i,j,u}$. ${\bf x}^{\rm ul}_i \!\in\! \bbC^{N_u}$ and ${\bf s}^{\rm ul}_i \!\in\! \bbC^{N_u}$ are the transmit signal and symbol vectors of the $N_u$ users in cell $i$, whose $u$th entries are $x^{\rm ul}_{i,u}$ and $s^{\rm ul}_{i,u}$, respectively. ${\pmb \Lambda}_i = {\rm diag}(\lambda_{i,1},\dots,\lambda_{i,N_u})$ collects the transmit power of the users in cell $i$, and $\bn^{\rm ul}_i \!\sim\! \mathcal{CN}({\bf 0}_{\!N_b},{\bf I}_{\!N_b})$ is the additive white Gaussian noise at BS$_i$.
We assume that  $\bs^{\rm ul}_i\!\sim\!\mathcal{CN}({\bf 0}_{\!N_u},{\bf I}_{\!N_u}), \forall i$.
\eqref{eq:r_ul} is merged in a compact form as
\begin{align}
    {\bf r}^{\rm ul}_i = {\bf H}_i{\pmb \Lambda}^{1/2}{\bf s}^{\rm ul} + {\bf n}_i^{\rm ul}
\end{align} 
where ${\bf H}_i \!\!=\!\! [{\bf H}_{i,1},\!..,\!{\bf H}_{i,N_c}] \!\in\!\! \bbC^{N_b\times N_cN_u}$\!, ${\pmb \Lambda} \!\!=\!\! {\rm blkdiag}({\pmb \Lambda}_{\!1},\!.., \!{\pmb \Lambda}_{\!N_c}\!) \! \\ \in\! \bbC^{N_cN_u\times N_cN_u}$, and ${\bf s}^{\rm ul} \!=\! [({\bf s}_1^{\rm ul})^{T},\!..,({\bf s}_{N_c}^{\rm ul})^{T}]^T \!\in\! \bbC^{N_cN_u}$.
% \begin{align}
%     &{\bf H}_i = [{\bf H}_{i,1},\dots,{\bf H}_{i,N_c}]\\
%     &{\pmb \Lambda} = {\rm diag}({\pmb \Lambda}_1,\dots, {\pmb \Lambda}_{N_c})\\
%     &{\bf s}^{T} = [{\bf s}_1^{T},\dots,{\bf s}_{N_c}^{T}].
% \end{align}

%We consider  that each ADC has $b$ quantization bits.
For analytical tractability, we adopt the AQNM~\cite{fletcher2007robust, orhan2015low} to have a linear approximation of a non-linear quantizer derived from a scalar MMSE quantizer.
% The AQNM is accurate enough in low and medium SNR ranges \cite{orhan2015low}.
Then the quantized signal vector can be given as~\cite{fletcher2007robust}
% EQUATION
\begin{align} 
     \cQ(\br_i)
    % & =\alpha {\bf r}_i + \bq_i^{\rm ul}\\
    % =  \alpha {\bf H}_i{\pmb \Lambda}^{1/2}{\bf s}^{\rm ul} + \alpha {\bf n}^{\rm ul}_i + \bq_i^{\rm ul}\\ 
    &\!\approx\!\alpha {\bf r}^{\rm ul}_i + \bq_i^{\rm ul}\\
    \label{eq:rq_ul}
    &\!=\! \alpha {\bf H}_{i,i}{\pmb \Lambda}^{1/2}_i{\bf s}_i^{\rm ul} \!+\!  \alpha {\textstyle\sum}_{j\neq i}^{N_c}{\bf H}_{i,j}{\pmb \Lambda}^{1/2}_j{\bf s}_j^{\rm ul} \!+\!  \alpha {\bf n}_i^{\rm ul} \!+\! \bq_i^{\rm ul}
\end{align} 
where $\mathcal{Q}(\cdot)$ is an entry-wise quantizer of the imaginary and real parts.
The quantizer gain $\alpha$ is a function of $b$, and defined as $\alpha \!=\! 1\!-\! \beta$, where $\beta\!=\! \frac{\mathbb{E}[|{r} - {r}_{{\rm q}}|^2]}{\mathbb{E}[|{r}|^2]}$ \cite{fletcher2007robust,fan2015uplink}.
$\beta$'s are quantified in Table 1 in \cite{fan2015uplink} for $b \leq 5$ assuming $\bs^{\rm ul}_i\!\sim\!\mathcal{CN}({\bf 0}_{N_u},{\bf I}_{N_u}), \forall i$. 
% Note that $b$ is the number of quantization bits for each real and imaginary part of each element in ${\bf r}_i$.
The quantization noise $\bq_i^{\rm ul}$ is uncorrelated with ${\bf r}_i$ and follows $\mathcal{CN}({\bf 0}_{N_b}, \mathbf{C}_{\bq_i^{\rm ul}\bq_i^{\rm ul}})$ with covariance of \cite{fletcher2007robust,orhan2015low}
% EQUATION
\begin{align}
   \label{eq:covariance_UL}
    \mathbf{C}_{\bq_i^{\rm ul}\bq_i^{\rm ul}} = \alpha \beta \,{\rm diag}\big({\bH_i}{\pmb \Lambda}{\bf H}_{i}^H + \bI_{N_b}\big).
\end{align}
The received signals are quantized and combined by ${\bf F}_i$ as
\begin{align}
    &{\bf y}^{\rm ul}_i  = {\bf F}_i^H \cQ(\br_i)
    %&= \!\alpha {\bf F}_i^H  {\bf H}_{i,i}{\pmb \Lambda}^{1\!/2}_i{\bf s}^{\rm ul}_i \!\!+\!  \alpha\!\! \sum_{j\neq i}^{N_c} \! {\bf F}_i^H  {\bf H}_{i,j}{\pmb \Lambda}^{1/2}_j{\bf s}^{\rm ul}_j  \!\!+\!  \alpha {\bf F}_i^H  {\bf n}^{\rm ul}_i \!+\! {\bf F}_i^H \!\bq_i^{\rm ul}. \nonumber
\end{align}
Noting that $\bff_{i,u}$ is the $u$th column of $\bF_i$, the combined signal for user $u$ in cell $i$ is written as
\begin{align}
    \nonumber
    &y^{\rm ul}_{i,u}  \approx \alpha \sqrt{\lambda_{i,u}} {\bf f}_{i,u}^H {\bf h}_{i,i,u} s^{\rm ul}_{i,u} \\
    &+\alpha {\textstyle\sum}_{(j,v) \neq (i, u)}^{(N_c, N_u)}\! \sqrt{\lambda_{j,v}} {\bf f}_{i,u}^H {\bf h}_{i,j,v} s^{\rm ul}_{j,v} 
    \!+\! \alpha  {\bf f}_{i,u}^H {\bf n}^{\rm ul}_{i} \!+\!  {\bf f}_{i,u}^H\bq_i^{\rm ul} 
\end{align}

%%%%%%%%%%%%%%%%%%%%%%%%
\subsection{Downlink System}
%%%%%%%%%%%%%%%%%%%%%%%%

% Now we consider a DL narrowband system.
%Similar to the UL quantized signals, 
The transmit signal vector quantized at low-resolution DACs of  BS$_i$ with a precoder ${\bf W}_i \in \bbC^{N_b\times N_u}$ is expressed as ${\bf x}^{\rm dl}_i \!=\! \alpha {\bf W}_i{\bf s}^{\rm dl}_i \!+\! \bq^{\rm dl}_i \in \bbC^{N_b}$ with the AQNM, where ${\bf s}^{\rm dl}_i \!\sim\! \mathcal{CN}({\bf 0}_{N_u},\!{\bf I}_{N_u}\!)$ denotes the transmit symbols dedicated to the $N_u$ users in cell $i$, and $\bq^{\rm dl}_i \!\in\! \bbC^{N_b}$ is a quantization noise vector with a covariance of \cite{dai2019achievable}
\begin{align}
    {\bf C}_{\bq^{\rm dl}_i\bq^{\rm dl}_i} = \alpha \beta {\rm diag}({\bf W}_i{\bf W}_i^H).
\end{align}
For the quantization, the same assumptions as the UL formulation are used and $\alpha$ is also identical to the one in the UL.
Under TDD assumption, the channel between BS$_j$ and user $u$ in cell $i$ is $\bh^H_{j,i,u}$. 
The received signal at user $u$ in cell $i$ is
\begin{align}
    \nonumber
    y^{\rm dl}_{i,u} \!=\!\ &\alpha {\bf h}_{i,i,u}^H \!{\bf w}_{i,u} {s}_{i,u}^{\rm dl} \!+\! \alpha\!\!\!\!\!\!\!\!\! \sum_{(j,v)\neq (i,u)}^{(N_c,N_u)} \!\!\!\!\!\!\!\!{\bf h}_{j,i,v}^H  \!{\bf w}_{j,v} {s}^{\rm dl}_{j,v} \!+\!\!\! \sum_{j=1}^{N_c} {\bf h}_{j,i,u}^H \bq_j^{\rm dl} \!+\!  n^{\rm dl}_{i,u} \nonumber
    % y^{\rm dl}_{i,u} = \alpha {\bf h}_{i,i,u}^H {\bf w}_{i,u} {s}_{i,u}^{\rm dl} + \alpha \sum_{v \neq u}^{N_u}{\bf h}^H_{i,i,u} {\bf w}_{i,v} {s}^{\rm dl}_{i,v} + \alpha \sum_{\substack{ j \neq i\\ v}}^{N_c,N_u} {\bf h}_{j,i,v}^H  {\bf w}_{j,v} {s}^{\rm dl}_{j,v} + \sum_{j=1}^{N_c} {\bf h}_{j,i,u}^H \bq_j +  n^{\rm dl}_{i,u}.
\end{align}
where $\bw_{i,u}$ is the $u$th column of $\bW_i$ and $n^{\rm dl}_{i,u}\sim\mathcal{CN}(0, 1)$.

%%%%%%%%%%%%%%%%%%%%%%%%
\section{Joint Beamforming and Power Allocation}
%%%%%%%%%%%%%%%%%%%%%%%%
In this section, we first write the UL and DL total transmit power minimization problems with SINR constraints, and then propose an algorithm that solve the problems.
%We assume that the problems are feasible.
The UL problem is built up to minimize the transmit power of the users in $N_c$ cells with an individual SINR constraint as
\begin{gather}
    \label{eq:problem_ul}
    \cP1:\min_{{\bf f}_{i,u}, \lambda_{i,u}, \forall i,u} {\textstyle\sum}_{i,u} \lambda_{i,u} \\ 
    \label{eq:problem_ul1}
    \text{s.t.}\; \max_{{\bf f}_{i,u}} \Gamma^{\rm ul}_{i,u} \geq \gamma_{i,u},\forall \, i,u
\end{gather}
where $\Gamma^{\rm ul}_{i,u}$ is the UL SINR of user $u$ in cell $i$ derived as
% The SINR for the user $u$ in the cell $i$ is calculated as 
\begin{align}
    \label{eq:sinr_ul}
    &\Gamma^{\rm ul}_{i,u}= \\
    & \frac{\alpha^2 \lambda_{i,u} |{\bff}_{i,u}^H {\bf h}_{i,i,u}|^2}{\alpha^2 \sum_{(j,v) \neq (i,u) }^{(N_c,N_u)} \lambda_{j,v} |{\bff}_{i,u}^H {\bf h}_{i,j,v}|^2  + {\bff}_{i,u}^H {\bf C}_{\bq^{\rm ul}_i\bq^{\rm ul}_i}{\bff}_{i,u} +  \alpha^2 \|{\bff}_{i,u}\|^2}. \nonumber
\end{align}
Compared with perfect quantization, the SINR derivation is intertwined with quantizer error in the middle of the denominator, i.e., ${\bff}_{i,u}^H {\bf C}_{\bq^{\rm ul}_i\bq^{\rm ul}_i}{\bff}_{i,u}$, whose magnitude increases as the quantization bits decrease due to the $\alpha\beta$ factor in \eqref{eq:covariance_UL}.
%which is a function of the channel and the transmit power $\lambda_{i,u}$ as well as the combiner $\bff_{i,u}$. 

We compose the DL minimum transmit power problem of the BSs with a user-wise target SINR constraint as
 \begin{gather}
    \label{eq:problem_dl}
    \cP2:\min_{{\bf w}_{i,u}, \forall i,u}   \alpha  {\textstyle\sum}_{i,u}   {\bf w}_{i,u}^H{\bf w}_{i,u}\\
    \label{eq:problem_dl1}
    \text{s.t.}\  \Gamma^{\rm dl}_{i,u} \geq \gamma_{i,u},\forall \, i,u
\end{gather}
where
\begin{align}
    \label{eq:sinr_dl}
    &\Gamma^{\rm dl}_{i,u} =\\
    & \frac{\alpha^2 |{\bf w}_{i,u}^H {\bf h}_{i,i,u}|^2} { \alpha^2 \sum_{(j,v) \neq (i,u)}^{(N_c,N_u)}|{\bf w}_{j,v}^H {\bf h}_{j,i,u}|^2  + \sum_{j=1}^{N_c}{\bf h}_{j,i,u}^H {\bf C}_{\bq^{\rm dl}_j\bq^{\rm dl}_j} {\bf h}_{j,i,u}\!+\!1}. \nonumber
    % \Gamma^{\rm dl}_{i,u} = \frac{\alpha^2 |{\bf w}_{i,u}^H {\bf h}_{i,i,u}|^2} { \alpha^2 \sum_{v \neq u} {| {\bf w}_{i,v}^H {\bf h}_{i,i,u}|^2} + \alpha^2 \sum_{\substack{j \neq i\\v}}|{\bf w}_{j,v}^H {\bf h}_{j,i,u}|^2  + \sum_{j}{\bf h}_{j,i,u}^H {\bf C}_{\bq^{\rm dl}_j\bq^{\rm dl}_j} {\bf h}_{j,i,u} + 1}.
\end{align}
% The solution of $\cP2$ also has to incorporate the coarse quantization effect, i.e., ${\bf h}_{j,i,u}^H {\bf C}_{\bq^{\rm dl}_j\bq^{\rm dl}_j}{\bf h}_{j,i,u}$.

%quantization noise covariance $\bC_{\bq^{\rm dl}_j\bq^{\rm dl}_j}$ as it is a function of $\bW_j$ and involved with channels ${\bf h}_{j,i,u}$.

%%%%%%%%%%%%%%%%%%%%%%%%%%%%%%%%%%%
\subsection{Uplink and Downlink Duality}
%%%%%%%%%%%%%%%%%%%%%%%%%%%%%%%%%%%

 By integrating the quantization error terms, we broaden the duality of the UL and DL power minimization problems for infinite-resolution quantizers \cite{dahrouj2010coordinated} to low-resolution quantizers. 
% In high-resolution ADC/DAC systems, it was proven that a strong duality holds between the UL and DL transmit power minimization problems.
% However, it is not known in low-resolution ADC/DAC systems due to the non-negligible quantizatiton error, which is a function of transmit power and beamformer as shown in \eqref{eq:sinr_ul} and \eqref{eq:sinr_dl}.
% In this subsection, we prove that the duality holds between the UL and the DL power optimization problems under the coarse quantization systems. 
%Using the duality, we introduce an iterative algorithm based on the fixed-point iteration \cite{yates1995framework} to solve both the UL and DL problems and further prove optimality and convergence.

\begin{theorem}[Duality]
    \label{thm:duality}
    The uplink transmit power minimization problem $\cP1$ in \eqref{eq:problem_ul}-\eqref{eq:problem_ul1} equals to the Lagrangian dual of the downlink transmit minimization problem $\cP2$ in \eqref{eq:problem_dl}-\eqref{eq:problem_dl1}.
\begin{proof}
    We use the MMSE combiners that maximize the SINR to simplify the constraints.
    Let ${\bf z}_{i,u}$ be the interference-plus-noise term of the quantized signal in \eqref{eq:rq_ul} with covariance of
    \begin{align}
        {\bf C}_{{\bf z}_{i,u}\!{\bf z}_{i,u}} 
         \!=\! \alpha^2 \!\! &\sum_{(j,v)\neq (i,u)} \!\!\!\!\!\!\!\lambda_{j,v} \!{\bf h}_{i,j,v}{\bf h}_{i,j,v}^H  \nonumber 
         \!\!+\! \alpha {\bf I}_{N_b} \!\!+\! \alpha\beta{\rm diag}(\!{\bf H}_i {\pmb \Lambda} {\bf H}_i^H\!).
    \end{align}
    Then, the linear MMSE equalizer ${\bf f}_{i,u}$ can be expressed as 
    \begin{align}
        \label{eq:MMSE}
        {\bf f}_{i,u} = {\bf C}_{{\bf z}_{i,u}{\bf z}_{i,u}}^{-1} {\bf h}_{i,i,u}.
    \end{align}
    Applying \eqref{eq:MMSE} to the UL SINR in \eqref{eq:sinr_ul}, the constraints in $\cP1$ are simplified as  $\alpha^2 \lambda_{i,u} {\bf h}_{i,i,u}^H {\bf C}_{{\bf z}_{i,u}}^{-1} {\bf h}_{i,i,u} \geq \gamma_{i,u}$.
    We then multiply both sides with $\bh_{i,i,u}^H\bh_{i,i,u}$ and rearrange as
    \begin{align}
        % \nonumber
        % &\alpha^2 \lambda_{i,u} {\bf h}_{i,i,u}^H{\bf h}_{i,i,u} {\bf h}_{i,i,u}^H {\bf C}_{{\bf z}_{i,u}{\bf z}_{i,u}}^{-1} {\bf h}_{i,i,u} \geq \gamma_{i,u}{\bf h}_{i,i,u}^H{\bf h}_{i,i,u}\\
        \label{eq:duality_pf}
        &{\bf h}_{i,i,u}^H(\alpha^2 \lambda_{i,u} {\bf h}_{i,i,u} {\bf h}_{i,i,u}^H {\bf C}_{{\bf z}_{i,u}{\bf z}_{i,u}}^{-1}\!\!-\gamma_{i,u} {\bf I}_{N_b}) {\bf h}_{i,i,u} \geq 0.
    %   \\
    %   \frac{\alpha}{\gamma_{i,u}} \lambda_{i,u} {\bf h}_{i,i,u} {\bf h}_{i,i,u}^H - \frac{1}{\alpha}{\bf C}_{{\bf z}_{i,u}} &\succeq 0
    \end{align}
    Here, \eqref{eq:duality_pf} implies that $\alpha^2 \lambda_{i,u} {\bf h}_{i,i,u} {\bf h}_{i,i,u}^H \!{\bf C}_{{\bf z}_{i,u}{\bf z}_{i,u}}^{-1}\!-\gamma_{i,u} {\bf I}_{N_b}$ needs to be a positive semidefinte matrix.
    Rearranging this condition, we can rewrite $\cP1$ as 
    % \begin{align}
    %     \label{eq:problem_ul_simple}
    %     &\min_{\lambda_{i,u}} \sum_{i,u} \lambda_{i,u} \\
    %     &\text{s.t. } {\bf K}_{i}(\pmb\Lambda) \preceq \alpha \bigg(1 + \frac{1}{\gamma_{i,u}}\bigg)\lambda_{i,u}  {\bf h}_{i,i,u}{\bf h}_{i,i,u}^H, \forall i,u. 
    % \end{align}
        \begin{gather}
        \label{eq:problem_ul_simple}
        \min_{\lambda_{i,u}} {\textstyle\sum}_{i,u} \lambda_{i,u} \\
        \label{eq:problem_ul_simple1}
        \text{s.t. } {\bf K}_{i}(\pmb\Lambda) \preceq \alpha \bigg(1+ \frac{1}{\gamma_{i,u}}\bigg)\lambda_{i,u}  {\bf h}_{i,i,u}{\bf h}_{i,i,u}^H,
    \end{gather}
    for all $i, u$ where 
    \begin{align}
        \nonumber
        {\bf K}_{i}(\pmb \Lambda) \!=\! {\bf I}_{N_b}  \!+\! \alpha {\textstyle\sum}_{j,v} \lambda_{j,v}{\bf h}_{i,j,v}{\bf h}_{i,j,v}^H \!+\! \beta {\rm diag}\big({\bf H}_i {\pmb \Lambda} {\bf H}_i^H\big).
    \end{align}
    
    We show the duality between $\cP1$ and $\cP2$ by handling the quantizer error and by proving that the problem in
    \eqref{eq:problem_ul_simple}-\eqref{eq:problem_ul_simple1} is equal to the Lagrangian dual of $\cP2$. The Lagrangian is given in \eqref{eq:lagrangian_pf1} in the next page
    % \begin{align} \label{eq:lagrangian}
    %     \nonumber
    %     \mathcal{L}(&{\bf w}_{i,u}, \mu_{i,u}) \!=\!  {\textstyle\sum}_{i,u} \alpha {\bf w}_{i,u}^H{\bf w}_{i,u} \!-\! {\textstyle\sum}_{i,u} \mu_{i,u} \bigg( \alpha^2 \frac{|{\bf w}^H_{i,u} {\bf h}_{i,i,u}|^2}{\gamma_{i,u}} \\
    %     &\;\;- \alpha^2 {\textstyle\sum}_{v \neq u} {| {\bf w}_{i,v}^H {\bf h}_{i,i,u}|^2}  - \alpha^2 {\textstyle\sum}_{\substack{j \neq i\\v}}|{\bf w}_{j,v}^H {\bf h}_{j,i,u}|^2  \nonumber \\
    %     &\;\;+ \alpha (1-\alpha) {\textstyle\sum}_{j}{\bf h}_{j,i,u}^H  {\rm diag}\big({\bf W}_j{\bf W}_j^H\big) {\bf h}_{j,i,u} + 1\bigg)\\
    %     \nonumber
    %     & = {\textstyle\sum}_{i,u}\mu_{i,u} + \alpha {\textstyle\sum}_{i,u} {\bf w}_{i,u}^H \bigg( \alpha {\textstyle\sum}_{j,v} \mu_{j,v}{\bf h}_{i,j,v}{\bf h}_{i,j,v}^H \\
    %     &- \alpha \bigg(1 + \frac{1}{\gamma_{i,u}}\bigg)\mu_{i,u}  {\bf h}_{i,i,u}{\bf h}_{i,i,u}^H  +{\bf I}_{N_b} \bigg){\bf w}_{i,u} \nonumber \\
    %      &+ \alpha (1-\alpha)  {\textstyle\sum}_{i,u}   \mu_{i,u}{\textstyle\sum}_{j}{\bf h}_{j,i,u}^H  {\rm diag}({\bf W}_j{\bf W}_j^H) {\bf h}_{j,i,u}.  \label{eq:lagrangian_pf1}
    % \end{align}
    \begin{figure*}
    \center
    \begin{align} 
    % \label{eq:lagrangian}
        \nonumber
        &\mathcal{L}({\bf w}_{i,u}, \mu_{i,u}) \\
         \nonumber
        &=\!  \sum_{i,u}\! \alpha {\bf w}_{\!i,u}^H{\bf w}_{\!i,u} \!-\! \sum_{i,u} \mu_{i,u} \!\bigg[ \alpha^2  \frac{|{\bf w}^H_{\!i,u} {\bf h}_{i,i,u}|^2}{\gamma_{i,u}} \!-\! \alpha^2 \!\sum_{v \neq u} {| {\bf w}_{\!i,v}^H {\bf h}_{i,i,u}|^2}  \!-\! \alpha^2 \!\sum_{\substack{j \neq i\\v}}|{\bf w}_{\!j,v}^H \!{\bf h}_{j,i,u}|^2  \!+\! \alpha\beta \sum_{j}{\bf h}_{j,i,u}^H  {\rm diag}\big({\bf W}_{\!\!j}{\bf W}_{\!\!j}^{\!H}\big) {\bf h}_{j,i,u} \!+\! 1\bigg] \\
        & =\! \sum_{i,u}\!\mu_{i,u} \!\!+\! \alpha \!\sum_{i,u} \!{\bf w}_{\!i,u}^{\!H} \bigg[ \alpha\! \sum_{j,v} \!\mu_{\!j,v}{\bf h}_{i,j,v}{\bf h}_{i,j,v}^{\!H} 
        \!\!-\! \alpha\!\! \left(\!\!1\!+\!\frac{1}{\gamma_{i,u}}\!\!\right)\!\!\mu_{i,u} \!{\bf h}_{i,i,u}{\bf h}_{i,i,u}^{\!H}  \!\!+\!{\bf I}_{\!N_{\!b}} \!\bigg]\!{\bf w}_{\!i,u} 
         \!\!+\! \alpha \beta\!  \sum_{i,u} \!\mu_{i,u}\!\!\sum_{j}\!{\bf h}_{j,i,u}^H  {\rm diag}(\!{\bf W}_{\!j}\!{\bf W}_{\!j}^{\!H}\!) {\bf h}_{j,i,u}.  \label{eq:lagrangian_pf1}
    \end{align}
    \hrulefill
    \end{figure*}
    where $\mu_{i,u}$'s are Lagrangian multipliers.
    We rewrite the quantization error term in \eqref{eq:lagrangian_pf1} to manage $\bW_{\!j}\!\bW_{\!j}^H$ inside the diagonal operator.
    Let $\bM_i \!\!=\!\! {\rm diag}(\mu_{i,1},\!..,\mu_{i,N_u})$ and ${\bM} \!=\! {\rm blkdiag}({\bM}_{1},\!.., \!{\bM}_{N_{c}}\!).$
    Rearranging the indices from $(i,u,j)$ to $(j,v,i)$ of $\sum_{i,u}\! \mu_{i,u}\!\sum_{j}\!{\bf h}_{j,i,u}^H  {\rm diag}(\!{\bf W}_{\!j}\!{\bf W}_{\!j}^H\!) {\bf h}_{j,i,u}$, we have
    \begin{align}
        \nonumber
        &{\textstyle\sum}_{j,v}^{(N_c,N_u)} \mu_{j,v}{\textstyle\sum}_{i}^{N_c}{\bf h}_{i,j,v}^H  {\rm diag}\big({\bf W}_i{\bf W}_i^H\big) {\bf h}_{i,j,v}\! \\
        % &=\!\sum_{j,v}^{N_c,N_u} \mu_{j,v} \sum_{i,n}^{N_c,N_b} |h_{i,j,v,n}|^2 \sum_u^{N_u}|w_{i,u,n}|^2\nonumber \\ 
        % \nonumber
        &=\!\!\!\!\sum_{i,u}^{N_c,N_u}\!\!\!\bw^H_{i,u}{\rm diag}\Big(\!\!\sum_{j,v}^{N_c,N_u} \!\!\!\mu_{j,v}|h_{i,j,v,1}|^2,\!..,\!\!\!\sum_{j,v}^{N_c,N_u} \!\!\!\mu_{j,v}|h_{i,j,v,N_b}|^2\!\Big) \bw_{i,u}\nonumber \\ 
        \label{eq:lagrangian_pf2} 
        &= \!{\textstyle\sum}_{i,u}^{(N_c,N_u)} {\bf w}_{i,u}^H {\rm diag}({\bf H}_i {\pmb \bM} {\bf H}_i^H){\bf w}_{i,u}, 
    \end{align}
    where $h_{i,j,v,n}$ and $w_{i,u,n}$ are the $n$th entries of $\bh_{i,j,v}$ and  $\bw_{i,u}$, respectively, and ${\bf H}_i \!=\! [{\bf H}_{i,1},\dots,{\bf H}_{i,N_c}]$ as previously defined.
    Since ${\bf w}_{i,u}$'s are factored out, we directly apply \eqref{eq:lagrangian_pf2} to \eqref{eq:lagrangian_pf1}, thereby rewriting the Lagrangian as
    \begin{align} 
        \nonumber
        &\mathcal{L}({\bf w}_{i,u}, \mu_{i,u}) \!=\! \!{\textstyle\sum}_{i,u}\mu_{i,u} \!\!+\! \alpha{\textstyle\sum}_{i,u} \! {\bf w}_{i,u}^H \!\bigg(
         \!\!\alpha {\textstyle\sum}_{j,v} \mu_{j,v}{\bf h}_{i,j,v}{\bf h}_{i,j,v}^H \! \\
        & \!\!\!\!\!+\!{\bf I}_{\!N_b} 
         \!\! -\! \alpha\! \bigg(\!\!1\! +\! \frac{1}{\gamma_{i,u}}\!\!\bigg)\!\mu_{i,u}\! {\bf h}_{i,i,u}\!{\bf h}_{i,i,u}^H \!\! +\! \beta  {\rm diag}\big({\bf H}_i {\bM} {\bf H}_i^{\!H}\big) \!\!\bigg)\!{\bf w}_{\!i,u}. \label{eq:lagrangian_pf3}
    \end{align}
    Let the dual objective function $g(\mu_{i,u}\!) \!=\! \min_{{\bf w}_{i,u}}\! \mathcal{L}({\bf w}_{i,u}, \mu_{i,u})$.  
    Not to have an unbounded solution, it is necessary to satisfy
    ${\bf I}_{N_b} - \alpha \Big(1 + \frac{1}{\gamma_{i,u}}\Big)\mu_{i,u}  {\bf h}_{i,i,u}{\bf h}_{i,i,u}^H + \alpha \sum_{j,v} \mu_{j,v}{\bf h}_{i,j,v}{\bf h}_{i,j,v}^H + \beta {\rm diag}({\bf H}_i {\bM} {\bf H}_i^H) \succeq 0$.
    Regrouping the expression, the Lagrangian dual of $\cP2$ in \eqref{eq:problem_dl}-\eqref{eq:problem_dl1} becomes equivalent to
    % \begin{align}\label{eq:lagrangian_pf3_2}
    %     &\max_{\mu_{i,u}} \sum_{i,u}^{N_c,N_u} \mu_{i,u} \\
    %     &{\text{s.t. }}\ {\bf K}_{i}(\bM) \succeq \alpha \bigg(1+\frac{1}{\gamma_{i,u}}\bigg)\mu_{i,u}  {\bf h}_{i,i,u}{\bf h}_{i,i,u}^H  , \forall, i,u
    % \end{align}
    \begin{gather}\label{eq:lagrangian_pf3_2}
        \max_{\mu_{i,u}} {\textstyle\sum}_{i,u} \mu_{i,u} \\ 
        \label{eq:lagrangian_pf3_2_1}
        {\text{s.t.}}\ {\bf K}_{i}(\bM) \succeq \alpha  \bigg(1+\frac{1}{\gamma_{i,u}}\bigg)\mu_{i,u} {\bf h}_{i,i,u}{\bf h}_{i,i,u}^H  ,
    \end{gather}
    for all $i, u$ where 
    \begin{equation}
        {\bf K}_{i}(\bM) \!=\! {\bf I}_{N_b}  \!+\! \alpha {\textstyle\sum}_{j,v} \mu_{j,v}{\bf h}_{i,j,v}{\bf h}_{i,j,v}^H + \beta{\rm diag}\big({\bf H}_i {\pmb \bM} {\bf H}_i^H\big). \nonumber
    \end{equation}
    We have the Lagrangian dual of $\cP2$ in \eqref{eq:lagrangian_pf3_2}-\eqref{eq:lagrangian_pf3_2_1} and the UL problem in \eqref{eq:problem_ul_simple}-\eqref{eq:problem_ul_simple1} with the flipped objectives and constraints.
    However, the optimal solutions of both cases are found at the active constraints.
    Since \eqref{eq:lagrangian_pf3_2}-\eqref{eq:lagrangian_pf3_2_1} and \eqref{eq:problem_ul_simple}-\eqref{eq:problem_ul_simple1} have the same objective value at active constraints,
    they become equivalent by replacing $\mu_{i,u}$'s in \eqref{eq:lagrangian_pf3_2}-\eqref{eq:lagrangian_pf3_2_1}  with $\lambda_{i,u}$'s.
\end{proof}
\end{theorem}
Noting that $\alpha \to 1$ as $b \to \infty$ , the results pave the way for a generalized understanding of the UL-DL duality derived in \cite{dahrouj2010coordinated} by extending it to any quantization resolution. To propose an algorithm that solves $\cP1$ and $\cP2$, and to prove its optimality, we show strong duality between $\cP1$ and $\cP2$.

\begin{corollary}
    \label{cor:strong_duality}
    No duality gap exists between $\cP2$ and its dual% problem.
    \begin{proof}
    We first show that $\cP2$ is an instance of a second-order cone programming.  
    % We can show this similarly to the proof presented in [Yonina].
    Let ${\bf W}$ be defined as ${\bf W} = [{\bf W}_1, \cdots, {\bf W}_{N_c}]$. The DL problem \eqref{eq:problem_dl} is reformulated as
    \begin{gather}
        \label{eq:strong_pf}
        \min_{{\bW},P_o} P_o \\ 
        \label{eq:strong_pf1}
        {\rm s.t.}\ \Gamma_{i,u}^{\rm dl} \geq \gamma_{i,u}, \quad \forall i,u\\ 
        \label{eq:strong_pf2}
        \quad \ {\rm Tr}\big({\bf W}^H{\bf W}\big) \leq P_o
    \end{gather}
    where $P_o$ is a positive slack variable.
    As noted in \cite{wiesel2005linear, yu2007transmitter}, we take a diagonal phase shifting on the right of the precoder of each cell as ${\bf W}_i{\rm diag}(e^{j\phi_{i,1}},\dots,e^{j\phi_{i,{N_u}}})$ for $i = 1,\cdots, N_c$.
    Therefore, we can design the precoder to be ${\bf w}_{i,u}^H{\bf h}_{i,i,u} \geq 0$, $\forall i,u$ without changing the objective nor the constraints.
    
    Using \eqref{eq:lagrangian_pf2}, we revise the quantization term in \eqref{eq:sinr_dl} as
    \begin{align}
        \label{eq:strong_pf3}
        \sum_j\!\bh_{\!j,i,u}^H\!\bC_{\!\bq^{\rm dl}_{j}\!\bq^{\rm dl}_{j}}\bh_{\!j,i,u}
        % &= \alpha(1-\alpha) \sum_j\bh_{j,i,u}^H{\rm diag}(\bW_j\bW_j^H)\bh_{j,i,u}
        \!=\!
        \alpha\beta\!\sum_{j,v}\!\bw_{\!j,v}^H{\rm diag}(\bh_{\!j,i,u}\bh_{\!j,i,u}^H\!)\bw_{\!j,v}.
    \end{align}
    Let $\bD_{j,i,u} \!=\! {\rm diag}(\bh_{j,i,u}\bh_{j,i,u}^H\!)$, $\bW_{\rm BD} \!=\! {\rm blkdiag}(\bW_1,\!..,\bW_{N_c}\!)$, and $\tilde{\bW}_{\rm BD} \!=\! {\rm blkdiag}((\bI_{N_b}\otimes\bW_1),\dots,(\bI_{N_b}\otimes\bW_{N_c}))$. 
    Using \eqref{eq:strong_pf3}, the SINR constraints in \eqref{eq:strong_pf1} becomes
    \begin{align}
        \nonumber
        \label{eq:strong_pf4}
        &\alpha^2 \bigg(1+\frac{1}{\gamma_{i,u}}\bigg) |{\bf w}_{i,u}^H {\bf h}_{i,i,u}|^2 \geq\\
        &\left\|
        \begin{matrix}
            \alpha\bW_{\rm BD}^H{\rm vec}(\bh_{1,i,u},\dots,\bh_{N_c,i,u}) \\
            \sqrt{\alpha(1-\alpha)}\tilde{\bW}_{\rm BD}^H{\rm vec}(\bD^{1/2}_{1,i,u},\dots,\bD^{1/2}_{N_c,i,u})\\
            1
        \end{matrix}
        \right\|^2, \forall \, i,u,
    \end{align}
    where ${\rm vec}()$ converts a matrix into a column vector. Because we force ${\bf w}_{i,u}^H {\bf h}_{i,u}$ to be non-negative, we take square root on both sides of \eqref{eq:strong_pf4}.
    In addition, \eqref{eq:strong_pf2} is reformulated as $ \|{\rm vec}({\bf W})\| \leq \sqrt{P_o}$. Thus,
    the problem in \eqref{eq:strong_pf}-\eqref{eq:strong_pf2} can be modified to the standard second order conic program \cite{wiesel2005linear}.
    
    Next, \eqref{eq:problem_dl} is strictly feasible because given a solution ${\bf W}$, it can be scaled by a factor of $c >1$ satisfying the constraints. 
    Thus, strong duality holds between \eqref{eq:problem_ul} and \eqref{eq:problem_dl}. 
    \end{proof}
\end{corollary}

%%%%%%%%%%%%%%%%%%%%%%%%%%%%%%%%%%%
\subsection{Distributed Iterative Algorithm}
\label{subsec:algorithm}
%%%%%%%%%%%%%%%%%%%%%%%%%%%%%%%%%%%

We specify solutions by using strong duality, and develop an iterative algorithm that finds the solutions for $\cP1$ and $\cP2$ concurrently. We further prove optimality and convergence.
\begin{corollary}
    \label{cor:solution}
    The optimal transmit power for the uplink total power minimization problem \eqref{eq:problem_ul} is derived as
    \begin{align}
        \label{eq:solution}
        \lambda_{i,u} = \frac{1}{\alpha \Big(1+\frac{1}{\gamma_{i,u}}\Big){\bf h}_{i,i,u}^H {\bf K}_i^{-1}({\pmb \Lambda}) {\bf h}_{i,i,u}}
    \end{align}
    where $ {\bf K}_{i}({\pmb \Lambda}) \!=\! {\bf I}_{N_b}  \!+\! \alpha \sum_{j,v} \lambda_{j,v}{\bf h}_{i,j,v}{\bf h}_{i,j,v}^H + \beta {\rm diag}({\bf H}_i {\pmb \Lambda} {\bf H}_i^H)$ 
    % \begin{align}
    %     {\bf K}_{i}({\pmb \Lambda}) = {\bf I}_{N_b}  + \alpha \sum_{j,v} \lambda_{j,v}{\bf h}_{i,j,v}{\bf h}_{i,j,v}^H + (1-\alpha) {\rm diag}({\bf H}_i {\pmb \Lambda} {\bf H}_i^H).
    % \end{align}
    with the linear MMSE equalizer given as
    \begin{align}
        \label{eq:MMSE2}
        {\bf f}_{i,u} \!\!=\!\!  \bigg[ \alpha^2 \!\!\!\!\!\sum_{(j,v)\neq (i,u)}\!\!\!\!\!\! \lambda_{j,v} {\bf h}_{i,\!j,\!v}{\bf h}_{i,\!j,\!v}^H \!\!+\! \alpha {\bf I}_{\!N_b}
        \!\!+\!\! \alpha\beta{\rm diag}(\!{\bf H}_i {\pmb \Lambda}\! {\bf H}_i^H\!)\!\bigg]^{\!-1}\!\!\!\!\!{\bf h}_{i,i,u}.
        % \label{eq:MMSE2}
        % {\bf f}_{i,u} =  \bigg[ \alpha^2 &\sum_{(j,v)\neq (i,u)} \lambda_{j,v} {\bf h}_{i,j,v}{\bf h}_{i,j,v}^H + \alpha {\bf I}_{N_b} \nonumber \\
        % &\qquad+ \alpha(1-\alpha){\rm diag}({\bf H}_i {\pmb \Lambda} {\bf H}_i^H)\bigg]^{-1}{\bf h}_{i,i,u}.
    \end{align}
\begin{proof}
    % Since the problem \eqref{eq:problem_dl} can be cast into SOCP and is strictly feasible, the variables in \eqref{eq:problem_ul} and \eqref{eq:problem_dl} are optimal if and only if its KKT conditions are satisfied.
    % Then, using the zero derivative condition, we can show that \eqref{eq:solution} is the optimal variables for \eqref{eq:problem_ul}.
    We use $\lambda_{i,u}$ instead of $\mu_{i,u}$ since we showed that they are equivalent.
    The derivative of the Lagrangian \eqref{eq:lagrangian_pf3} with respect to ${\bf w}_{i,u}$ is given as
    \begin{align} \label{eq:zero_derivative}
        &\frac{\partial \mathcal{L}({\bf w}_{i,u}, \lambda_{i,u})}{\partial {\bf w}_{i,u}} \!=\! 2\alpha \bigg({\bf I}_{N_b}  \!-\! \alpha \bigg(1 \!+\! \frac{1}{\gamma_{i,u}}\bigg)\lambda_{i,u}  {\bf h}_{i,i,u}{\bf h}_{i,i,u}^H \nonumber \\
        &\! + \!\alpha {\textstyle\sum}_{j,v} \lambda_{j,v}{\bf h}_{i,j,v}{\bf h}_{i,j,v}^H 
         \!+\! \beta {\rm diag}({\bf H}_i {\pmb \Lambda} {\bf H}_i^H) \bigg){\bf w}_{i,u}. 
    \end{align}
    Setting \eqref{eq:zero_derivative} to zero, we have \eqref{eq:solution}, i.e., the Lagrangian multiplier that meets the stationary condition. 
    Also, all the constraints in $\cP2$ are active at \eqref{eq:solution}, hereby satisfying the complementary slackness.
    % As the MMSE receiver maximizes the SINR, the combiner becomes \eqref{eq:MMSE2}
    Thus, \eqref{eq:solution} is the optimum of $\cP1$.
    % and this completes the proof.
\end{proof}
\end{corollary}

Given that we finally have the optimal transmit power, we can compute the optimal UL MMSE combiner in \eqref{eq:MMSE}. As a function of these factors, we design the optimal DL precoder.
\begin{corollary}
    \label{cor:DL_precoder}
    The optimal DL precoders are linearly proportional to the UL MMSE equalizer , i.e.,  $\bw_{i,u} = \sqrt{\tau_{i,u}}\bff_{i,u} \;\forall i, u$ and $\tau_{i,u}$ is derived as $\btau = \bSigma^{-1}{\bf 1}_{N_uN_c}$ where $\btau = [\btau_{1}^T, \btau_{2}^T, \!.., \btau_{N_c}^T ]^T$ with  $\btau_{i}^T = [\tau_{i,1}, \tau_{i,2}, \!.., \tau_{i,N_u}]^T$, and  
    \begin{equation}
    \bSigma = 
    \begin{pmatrix}
    \bSigma_{1,1} & \bSigma_{1,2} & \cdots & \bSigma_{1,N_c} \\
    \bSigma_{2,1} & \bSigma_{2,2} & \cdots & \bSigma_{2,N_c} \\
    \vdots  & \vdots  & \ddots & \vdots  \\
    \bSigma_{N_c,1} & \bSigma_{N_c,2} & \cdots & \bSigma_{N_c,N_c}
    \end{pmatrix},
    \end{equation}
    where each element of matrix $\bSigma_{i,j}\in\bbR^{N_u \times N_u}$ is given as
    \begin{align} 
        \label{eq:DLprecoder}
        &[\bSigma_{i,j}]_{u,v} = \\
        &\begin{cases}
        \frac{\alpha^2}{\gamma_{i,u}} |\bff_{i,u}^H {\bf h}_{i,i,u}|^2 \\
        \;\;\;\;\;\;- \alpha\beta\bff_{i,u}^H{\rm diag}(\bh_{i,i,u}\bh_{i,i,u}^H)\bff_{i,u} & \!\!\text{if } i=j \text{ and } u=v, \\
        -  \alpha^2 | \bff_{j,v}^H {\bf h}_{j,i,u}|^2 \\
        \;\;\;\;\;\;- \alpha\beta\bff_{j,v}^H{\rm diag}(\bh_{j,i,u}\bh_{j,i,u}^H)\bff_{j,v} & \!\!\text{otherwise.} \nonumber
        \end{cases}
    \end{align}
    %  {\color{red} ($\tau_{i,u} = ?$)}
    % {\color{red} (UL combiner: $\bff_{i,u}$, DL precoder: $\bw_{i,u}$ in our paper. Avoid using $\beta$ here since we have already been using it. Try $\tau$) instead}
    \begin{proof}
        Based on the Lagrangian dual, a global optimum appears when the constraints are active.
    By replacing $\bw_{i,u}$ in \eqref{eq:sinr_dl} with $\sqrt{\tau_{i,u}}\bff_{i,u}$, the constraints of the DL problem in \eqref{eq:problem_dl1} satisfy the following equality conditions: 
    \begin{align}
        \nonumber
        1&=\frac{\alpha^2}{\gamma_{i,u}} |{\bf w}_{i,u}^H {\bf h}_{i,i,u}|^2 -  \alpha^2 {\textstyle\sum}_{v \neq u} {| {\bf w}_{i,v}^H {\bf h}_{i,i,u}|^2} \\
        \nonumber
        &\qquad- \alpha^2 {\textstyle\sum}_{\substack{j \neq i\\v}}|{\bf w}_{j,v}^H {\bf h}_{j,i,u}|^2  - {\textstyle\sum}_{j}{\bf h}_{j,i,u}^H {\bf C}_{\bq^{\rm dl}_j\bq^{\rm dl}_j} {\bf h}_{j,i,u}  \\
        \nonumber
        &\stackrel{(a)}=  \frac{\alpha^2}{\gamma_{i,u}} |\bff_{i,u}^H {\bf h}_{i,i,u}|^2 \tau_{i,u} -  \alpha^2 {\textstyle\sum}_{(j,v) \neq (i,u)} {| \bff_{j,v}^H {\bf h}_{j,i,u}|^2} \tau_{j,v} \\
        \nonumber
        &\qquad- \alpha\beta{\textstyle\sum}_{j,v}\bff_{j,v}^H{\rm diag}(\bh_{j,i,u}\bh_{j,i,u}^H)\bff_{j,v}\tau_{j,v} , \quad \forall i, u, \nonumber
    \end{align} 
    where $(a)$ is from \eqref{eq:strong_pf3} and $\bw_{\!i,\!u} \!=\! \sqrt{\tau_{i,u}}\bff_{i,u}$. 
    We cascade the conditions $\forall i,u$ in a matrix form: $\bSigma\btau \!\!=\!\! {\bf 1}$ and $\btau \!\!=\!\! \bSigma^{-1}{\bf 1}$.
    \end{proof}
\end{corollary}

We devise the unified algorithm to solve both UL and DL problems. We first solve \eqref{eq:solution} in UL problem, however, the main drawback is that all the transmit powers engage in the computation of an individual transmit power. Therefore, we employ an iterative standard algorithm based on \eqref{eq:solution} \cite{yates1995framework,wiesel2005linear,dahrouj2010coordinated} to find the optimal UL solution.
% It is shown in \ref{col:strong_duality} that the strong duality holds, the optimal solution for the downlink problem \eqref{eq:problem_dl} is also the optimal solution for the uplink problem \eqref{eq:problem_ul}.
% Thus, in the following section, 
% we propose a fixed-point iterative algorithm which always converges to an unique fixed point that is the optimal transmit power.
% The algorithm is based on the idea of iterative function evaluation [Yonina][Dahrouj]. 
Let $\lambda_{i,u}^{(n)}$ be the result at $n$th iteration, and $\pmb \Lambda^{(n)}$ be a collection of $\lambda_{i,u}^{(n)}$'s.
The algorithm is described as follows:
\begin{description}
    \item[Step 1.] Initialize $\lambda_{i,u}^{(0)}$, $\forall i,u$.
    \item[Step 2.] Iteratively update $\lambda_{i,u}^{\!(\!n+1\!)}$ until converges using \eqref{eq:solution} as
        \begin{align}
            \nonumber
            \lambda_{i,u}^{(n+1)} = \frac{1}{\alpha \Big(1\!+\!\frac{1}{\gamma_{i,u}}\Big)\!{\bf h}_{i,i,u}^H {\bf K}_i^{-1}(\pmb \Lambda^{(n)}) {\bf h}_{i,i,u}},  \forall i,u.
        \end{align}
        % where 
        % \begin{align}
        %     {\bf K}_i^{(n)} =  {\bf I}  + \alpha \sum_{j,v} \lambda_{j,v}^{(n)}{\bf h}_{i,j,v}{\bf h}_{i,j,v}^H + (1-\alpha) {\rm diag}({\bf H}_i {\pmb \Lambda}^{(n)} {\bf H}_i^H).
        % \end{align}
    \item[Step 3.] Find the UL MMSE combiner ${\bf f}_{i,u}$ in \eqref{eq:MMSE2} with $\lambda_{i,u}$.
    \item[Step 4.] Compute the DL precoder $\bw_{i,u}$ based on Corollary~\ref{cor:DL_precoder}.
\end{description}
${\bf K}_{i}$ is a covariance matrix of received signals which may be estimated using local measurements at BS$_i$ \cite{dahrouj2010coordinated}; hence Step 2 does not entail explicit inter-cell channel knowledge. 
The individual weight $\tau_{i,u}$ that achieves the target SINR can be obtained using a per-user update \cite{foschini1993simple}, whose convergence is guaranteed \cite{yates1995framework}.
Each step of the algorithm evolves $\tau_{i,u}$ while assuming other $\tau_{i',u'}$'s are unchanged.
Thus, the proposed algorithm can be deployed in a distributed fashion.
\begin{corollary}[Convergence]
    \label{cor:convergence}
    For any initialization $\lambda_{i,u}^{(0)}$, $\forall i,u$, the proposed fixed-point iterative algorithm converges to an unique fixed point at which total transmit power is minimized.
    \begin{proof}
    We exploit the standard function \cite{yates1995framework}. 
    We need to show that $\mathcal{F}_{i,u}(\pmb \lambda)$ is a standard function which meets the following:
    \begin{itemize}
        \item (positivity) If $\lambda_{i,u}\geq 0$ $\forall i,u$, then $\mathcal{F}_{i,u}({\pmb \Lambda}) > 0$.
        \item (monotony) If $\!\lambda_{i,u}\! \geq\!\lambda_{i,u}' \forall i,u$, then $\!\mathcal{F}_{i,u}({\pmb \Lambda}) \!\geq\! \mathcal{F}_{i,u}({\pmb \Lambda}')$.  
        \item (scalability) For $\rho > 1$, $\rho \mathcal{F}_{i,u}({\pmb \Lambda}) > \mathcal{F}_{i,u}(\rho{\pmb \Lambda})$.
    \end{itemize}
    It can be shown that $\mathcal{F}_{i,u}(\pmb \Lambda^{(n)})$ satisfies these properties by carefully following the proof in Appendix II in \cite{wiesel2005linear}.
    \end{proof}
\end{corollary}

\section{Simulation Results}
%%%%%%%%%%%%%%%%%%%%%%%%%%%%%%%%%%

% Simulation Setting
We evaluate the derived results and the proposed quantization-aware iterative CoMP algorithm (Q-iCoMP) using our source code written in Matlab on
GitHub \cite{code}. 
As a benchmark, we test the quantization-aware per-cell iterative algorithm (Q-Percell) by adapting the per-cell algorithm in \cite{rashid1998transmit} to low-resolution converters. 
For Q-Percell, each BS first finds a solution based on the iterative process in \cite{rashid1998transmit} considering the inter-cell interference as fixed noise. 
Once the BSs find solutions for the considered noise power, the BSs update the noise power and iterate until the solutions converge.

We consider two networks with $N_c \in \{2,7\}$, which we call a light and dense network, respectively.
For $N_c=2$, two cells are next to each other. 
For $N_c=7$, the center cell is surrounded by the other six cells.  
% that $N_c$ cells are generated as 
Each BS is in the center of each hexagonal cell with $N_u$ users.
The distance between adjacent BSs is $2\ \rm km$ and a user is at least $100\ \rm m$ away from the BSs.
For small scale fading, we assume Rayleigh fading with a zero mean and unit variance. 
For large scale fading, we use the log-distance pathloss in \cite{erceg1999empirically}.
We consider $2.4\ \rm GHz$ carrier frequency, $10\  \rm MHz$ bandwidth, $8.7\  \rm dB$ lognormal shadowing variance, and $5\  \rm dB$ noise figure.
We assume the same target SINR $\gamma$ for all users over all cells.

% %%%%%%%%%%%%% <<FIGURE>> %%%%%%%%%%%%%
% \begin{figure}[!t]\centering
% \includegraphics[scale = 0.4]{cells.png}
% \caption{Network configuration for 2-cell and 7-cell networks} 
% \label{fig:cell}
% \end{figure}
% %%%%%%%%%%%%%%%%%%%%%%%%%%%%%%%%%%%%%%

%%%%%%%%%%%%% <<FIGURE>> %%%%%%%%%%%%%
\begin{figure}[!t]\centering
\includegraphics[width=0.96\columnwidth]{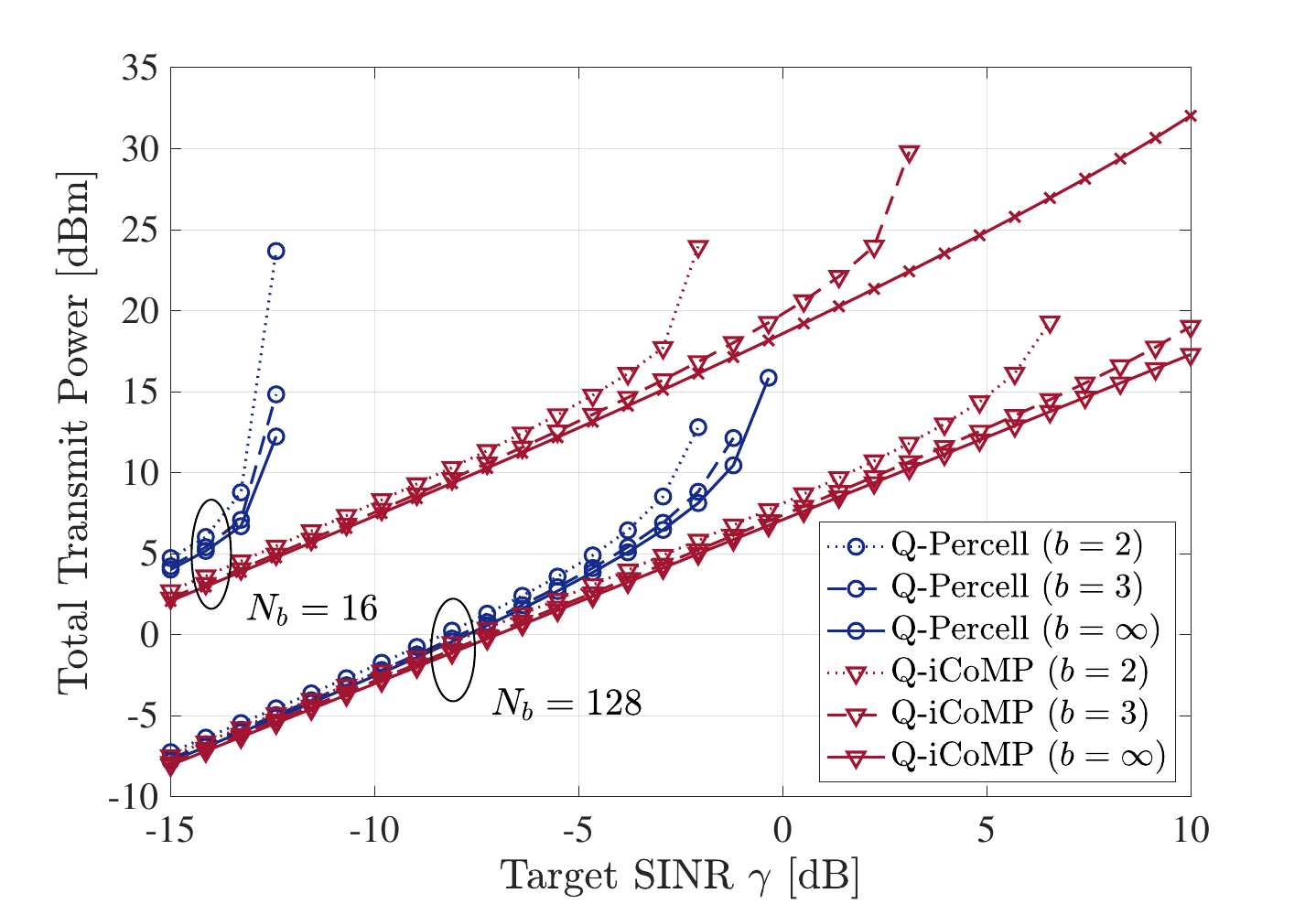}
\vspace{-1em}
\caption{Total transmit power versus the target SINR for $N_b \in \{16, 128\}$, $N_c = 7$ cells, and $N_u = 4$ users/cell.} 
\label{fig:txpower2}
\vspace{-1em}
\end{figure}
%%%%%%%%%%%%%%%%%%%%%%%%%%%%%%%%%%%%%%

Fig.~\ref{fig:txpower2} shows total transmit power for the target SINR. We vary the number of quantization bits and BS antennas, i.e., $b\in\{2,3,\infty\}$ and $N_b\in\{16,128\}$, for the dense network with $N_c \!=\! 7$ and $N_u \!=\! 4$.
For $N_b\!=\!16$, Q-Percell suffers from implausible power consumption even with infinite-resolution quantization at the low SINR requirement. Even though Q-iCoMP also shows divergence in total transmit power at medium to high target SINRs with a small number of quantization bits, Q-iCoMP shows a much slower rate of divergence compared to Q-Percell.
For $N_b \!=\!128$, Q-Percell has a similar trend as the case with $N_b\!=\!16$, showing  the divergence at medium SINR.
In contrast, Q-iCoMP reaches the target SINRs without divergence for $b \geq 3$ bits. On both  $N_b\in\{16,128\}$, Q-iCoMP achieves significant power gain over Q-Percell.  Increasing the number of antennas from $16$ to $128$ provides more than $10$ dB gain for each user SINR.
Accordingly, proper coordination is essential when deploying a massive antenna array with low-resolution quantizers.

%%%%%%%%%%%%% <<FIGURE>> %%%%%%%%%%%%%
\begin{figure}[!t]\centering
\includegraphics[width=0.96\columnwidth]{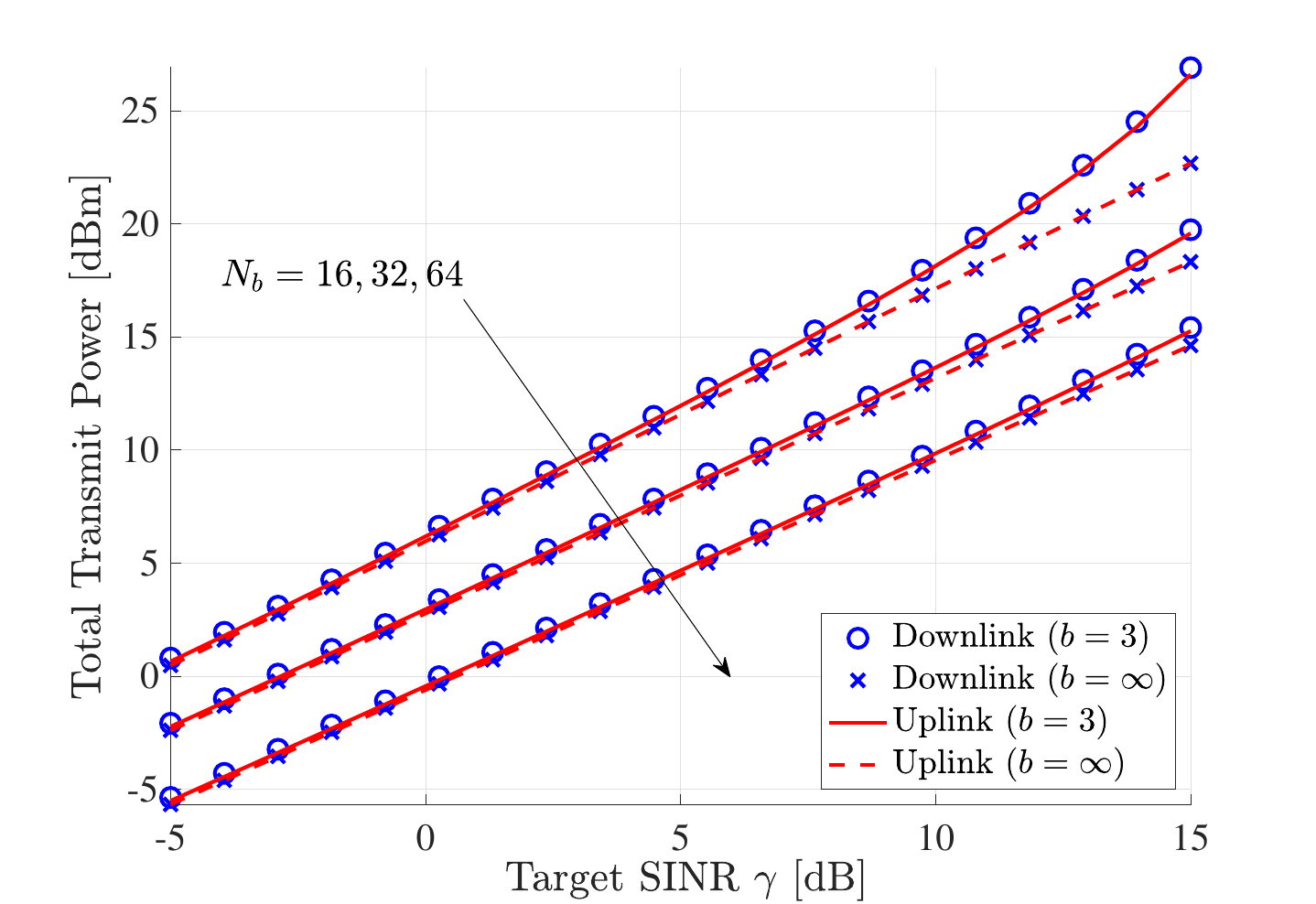}
\vspace{-1em}
\caption{Total transmit power versus the target SINR for the UL and DL cases with $N_b \in \{16, 32, 64\}$ BS antennas, $N_c = 3$ cells,  $N_u = 3$ users/cell, and $b \in \{3,\infty\}$ bits.} 
\label{fig:SD}
\vspace{-1em}
\end{figure}
%%%%%%%%%%%%%%%%%%%%%%%%%%%%%%%%%%%%%%

Fig.~\ref{fig:SD} shows the total transmit power  for both UL and DL networks with $N_b \in \{16, 32, 64\}$, $N_c = 3$,  $N_u = 3$, and $b \in \{3,\infty\}$. 
In all cases, the total transmit power for the DL and UL problems match, which validates strong duality, i.e., regardless of the number of quantization bits and the number of antennas, the considered UL and DL problems have optimal solutions that achieve the same minimum total transmit power.
% %%%%%%%%%%%%% <<FIGURE>> %%%%%%%%%%%%%
% \begin{figure}[!t]\centering
% \includegraphics[width=1\columnwidth]{antennas.eps}
% \vspace{-1em}
% \caption{Total transmit power versus the target SINR for $N_b \in \{16, 128\}$, $N_c = 7$ cells, and $N_u = 4$ users per cell.} 
% \label{fig:txpower2}
% \end{figure}
% %%%%%%%%%%%%%%%%%%%%%%%%%%%%%%%%%%%%%%

\begin{figure}[!t]
\centering
\includegraphics[width=1.01\columnwidth]{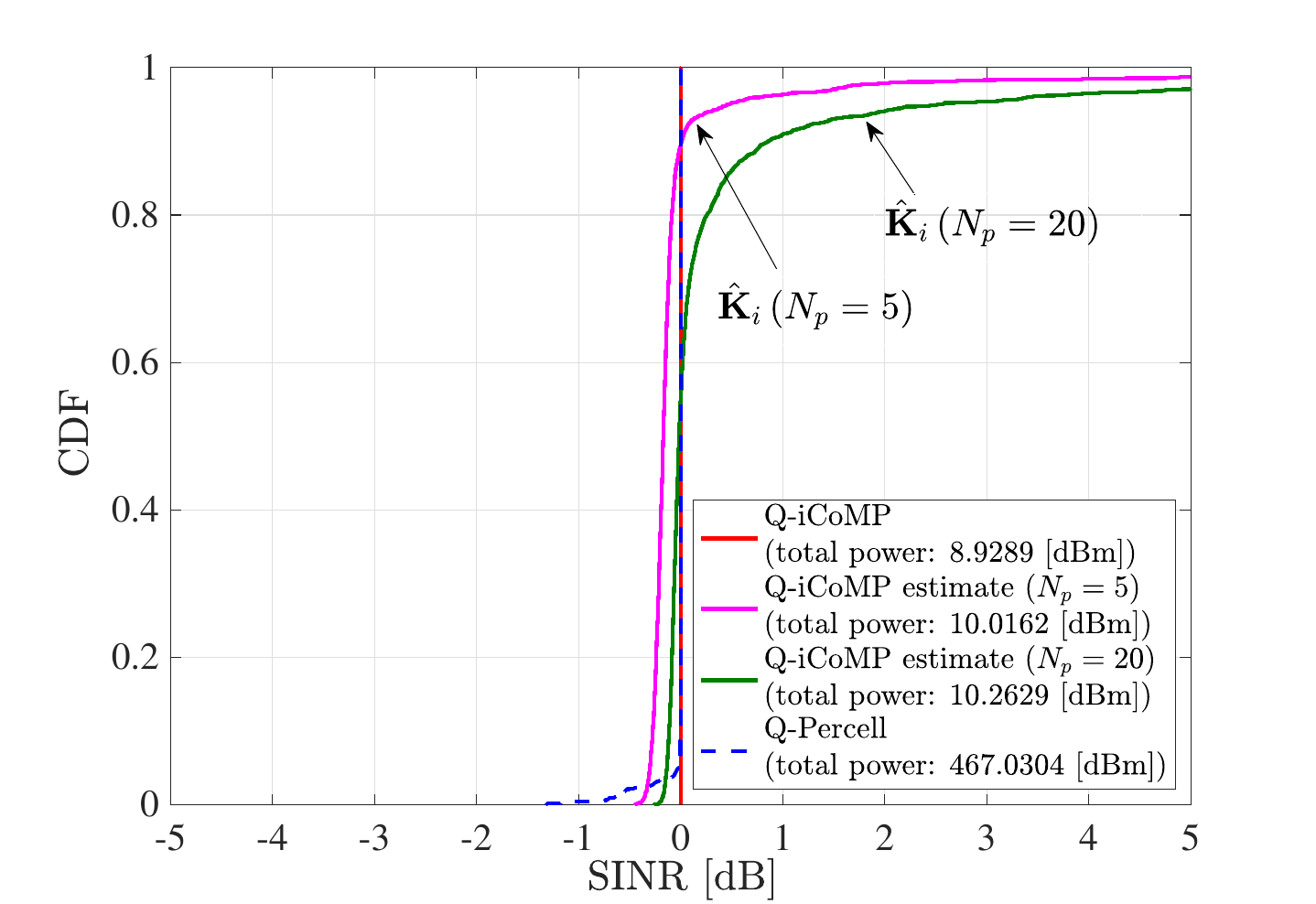}
\vspace{-2 em}
    \caption{CDFs of the SINRs of users in all cells for $\gamma = 0\  \rm dB$ target SINR, $b= 3$ bits, $N_b = 64$ BS antennas, $N_c = 7$ cells, and $N_u = 4$ users/cell.}
\label{fig:cdf}
\vspace{-1 em}
\end{figure}
%%%%%%%%%%%%%%%%%%%%%%%%%%%%%%%%%%

Fig.~\ref{fig:cdf} shows the cumulative density function (CDF) of the achieved SINR of all users for  $\gamma = 0\  \rm dB$, $b=3$, $N_b = 64$, $N_c = 7$, and $N_u = 4$. The total transmit power is annotated in the legend.
Q-iCoMP shows a clear spike at $0\ \rm dB$ with the least total transmit power.
Q-iCoMP properly controls the transmit power so that the achieved SINRs are no more and no less than what the system requires.
%This validates the performance of Q-iCoMP which provides an optimal solution for the UL and DL problems.
Although Q-Percell requires implausible power,
around $5\%$ of users cannot achieve the target SINR due to the lack of multicell coordination.
% We emphasize that the proposed Q-iCoMP method outperforms the conventional approach with more interferers.

Fig.~\ref{fig:cdf} also includes the result with the distributed method that estmates ${\bK}_i$ without explicit knowledge on inter-cell interference.
At $n$th iteration, the BSs have $\lambda_{i,u}^{(n-1)},\ \forall i,u$, in memory and each BS$_i$ demands the associated users to transmit $N_p$ pilots $\bs^{\rm ul}_i, \ \forall u,$ with the updated $\lambda_{i,u}^{(n-1)}$, and then averages out the local observations.
Distributed Q-iCoMP implementations with the estimation of $\bK_i$ with $N_p = 5$ and $N_p = 20$ are evaluated.
Both cases fulfill the target SINR with less than $-0.5$ dB deviation which decreases as $N_p$ increases.
Using 20 pilots further achieves the target SINR with higher probability compared to the estimation with 5 pilots.

% In Fig.~\ref{fig:txpower2}, the network with $N_c = 7$ and $N_u = 4$ is considered for the different $b$ and $N_b$.
% For $N_b=16$, Q-Percell algorithm is almost infeasible and Q-iCoMP algorithm also shows divergence in the total transmit power at the medium to high target SINRs with a small number of quantization bits.
% Increasing the number of BS antennas from $16$ to $128$ provides more than $10$ dB SINR gain.
% Accordingly, for $N_b =128$ which is considered as the massive MIMO system, Q-iCoMP algorithm achieves the target SINRs for all users without divergence even with $b = 3$, whereas Q-Percell algorithm still suffers from excessive power consumption in the medium to high target SINR range.
% Therefore, in massive MIMO systems, the coordinated joint BF and PA can provide reliable and power-efficient communications even with a small number of quantization bits, thereby achieving spectrum- and energy-efficient communications.

%%%%%%%%%%%%%%%%%%%%%%%%%%%%%%%%%%
\section{Conclusion}
%%%%%%%%%%%%%%%%%%%%%%%%%%%%%%%%%%

% Problem statement
This paper investigated the CoMP BF and PA for a multicell network with low-resolution ADCs and DACs.
% Duality
Incorporating the effect of non-trivial quantization error, we derived strong duality between the UL and DL total transmit power minimization problems under  user SINR constraints using low-resolution data converters.
% Iterative Algorithm
Using strong duality, we developed an iterative algorithm to find a fixed point solution that optimizes the UL and DL problems with the coarse quantization.
The proposed algorithm determines optimal solutions in a distributed fashion without explicit out-of-cell channel information. 
% Simulations
In simulation, the proposed iterative design is more effective than the conventional approach in terms of the total power consumption and achieved SINR. 
The performance gain increases the number of antennas and cells.
Our distributed coordination method works particularly well for a massive number of antennas with low-resolution ADCs and DACs.
For a more thorough discussion of our results, please see \cite{choi2020quantized}.

\bibliographystyle{IEEEtran}
\bibliography{CoMP_ADCs.bib}

\end{document}